\documentclass[aps,prb,reprint,longbibliography,superscriptaddress]{revtex4-2}
\usepackage{mathrsfs}
\usepackage{amsmath,gensymb}
\usepackage{amsfonts}
\usepackage{amssymb}
\usepackage{amsthm}
\usepackage{graphicx}
\usepackage{natbib}
\usepackage{color}
\usepackage{hyperref}
\usepackage{bm}
\usepackage[caption=false]{subfig}
\usepackage{verbatim}

\newcommand{\lina}[1]{{\color{black} #1}}

\newcommand{\prlsection}[1]{\textit{#1}.\kern0.05em---\kern0.05em\ignorespaces}

\newcommand{\ve}{\boldsymbol}

\newcommand{\ak}[1]{{\color{black} #1}}

\begin{document}
\title{\ak{Complete $T_c$ suppression} and N\'eel triplets-mediated exchange in antiferromagnet-superconductor-antiferromagnet trilayers}

\author{Lina Johnsen Kamra}
%\email{lina.g.johnsen@ntnu.no}
\thanks{These two authors contributed equally.}
\affiliation{Center for Quantum Spintronics, Department of Physics, Norwegian University of Science and Technology, NO-7491 Trondheim, Norway}
\affiliation{Condensed Matter Physics Center (IFIMAC) and Departamento de F\'{i}sica Te\'{o}rica de la Materia Condensada, Universidad Aut\'{o}noma de Madrid, E-28049 Madrid, Spain}

\author{Simran Chourasia}
\thanks{These two authors contributed equally.}
\affiliation{Condensed Matter Physics Center (IFIMAC) and Departamento de F\'{i}sica Te\'{o}rica de la Materia Condensada, Universidad Aut\'{o}noma de Madrid, E-28049 Madrid, Spain}

\author{G.~A.~Bobkov}
\affiliation{Moscow Institute of Physics and Technology, Dolgoprudny, 141700 Moscow, Russia}

\author{V.~M.~Gordeeva}
\affiliation{Moscow Institute of Physics and Technology, Dolgoprudny, 141700 Moscow, Russia}

\author{I.~V.~Bobkova}
\affiliation{Moscow Institute of Physics and Technology, Dolgoprudny, 141700 Moscow, Russia}
\affiliation{National Research University Higher School of Economics, Moscow, 101000 Russia}

\author{Akashdeep Kamra}
\affiliation{Condensed Matter Physics Center (IFIMAC) and Departamento de F\'{i}sica Te\'{o}rica de la Materia Condensada, Universidad Aut\'{o}noma de Madrid, E-28049 Madrid, Spain}

\begin{abstract}
An antiferromagnetic insulator (AFI) bearing a compensated interface to an adjacent conventional superconductor (S) has recently been predicted to generate N\'eel triplet Cooper pairs, whose amplitude alternates sign in space. Here, we theoretically demonstrate that such N\'eel triplets enable control of the superconducting critical temperature in an S layer via the angle between the N\'eel vectors of two enclosing AFI layers. This angle dependence changes sign with the number of S monolayers providing a distinct signature of the N\'eel triplets. Furthermore, we show that the latter mediate a similarly distinct exchange interaction between the two AFIs' N\'eel vectors.
\end{abstract}

\maketitle

%-----------------------------------------Intro------------------------------------------------------- %
\section{Introduction}
Hybrids comprising a conventional spin-singlet superconductor (S) and one or more magnetic layers realize unconventional superconductivity and Cooper pairs thereby enabling intriguing physics and \lina{potential applications}~\cite{Sigrist1991,Buzdin2005,Bergeret2005,Tanaka2012,Eschrig2015,Linder2015,Bergeret2018}. The central role of magnets in this engineering of superconductivity is to induce a spin-splitting field which generates spin-triplet Cooper pairs from their spin-singlet counterparts available in S~\cite{Maki1964,Clogston1962,Chandrasekhar1962}. This also reduces the \lina{superconducting} critical temperature $T_c$. A canonical structure sandwiches a thin S layer between two ferromagnet (F) layers and enables a control over the $T_c$ via the relative angle $\alpha$ between the two F magnetizations~\cite{Degennes1966,Deutscher1969,Tagirov1999,Kulic2000,Gu2002,Westerholt2005,Moraru2006,Li2013,Gu2015}. The dominant effect in this trilayer is the addition (cancellation) of spin-splitting fields from the two Fs when their magnetizations are parallel (antiparallel) resulting in the smallest (largest) $T_c$ out of all F configurations. This has been exploited to switch the S to its normal resistive state by controlling $\alpha$ via an applied magnetic field. This in turn admits a change of resistance from zero to a nonzero value i.e., an infinite magnetoresistance~\cite{Tagirov1999,Li2013,Gu2015}. %The same phenomenon also enables a superconducting spin switch and Cooper pairs-mediated coupling between the two F magnetizations.

The dipolar stray fields and GHz frequency magnons in F are parasitic detrimental influences in these devices. Employing antiferromagnets (AFs) could significantly reduce these problems due to their \lina{zero} net magnetization and higher magnon frequencies~\cite{Gomonay2014,Baltz2016,Jungwirth2016,Kamra2018}. Furthermore, their two or more sublattices admit intriguing phenomena that bring along entirely novel functionalities~\cite{Libor2018}. However, early experiments with metallic AFs found no influence on an adjacent S, attributing this to their lack of net magnetization and thus spin-splitting field~\cite{Werthammer1966}. Subsequently, there have been theoretical predictions of spin-dependent transport at such \lina{AF--S} interfaces~\cite{Bobkova2005,Andersen2005,Andersen2006,Jakobsen2020}. More recent experiments find a strong effect of the AF on an adjacent S layer~\cite{Hubener2002,Bell2003,Wu2013,Seeger2021,Mani2009,Mani2015}. An AF or antiferromagnetic insulator (AFI) bearing an uncompensated interface to the adjacent S has recently been shown to induce spin-splitting~\cite{Kamra2018}, which contributes to influencing the S. 

\begin{figure}[tb]
	\centering
	\includegraphics[width=85mm]{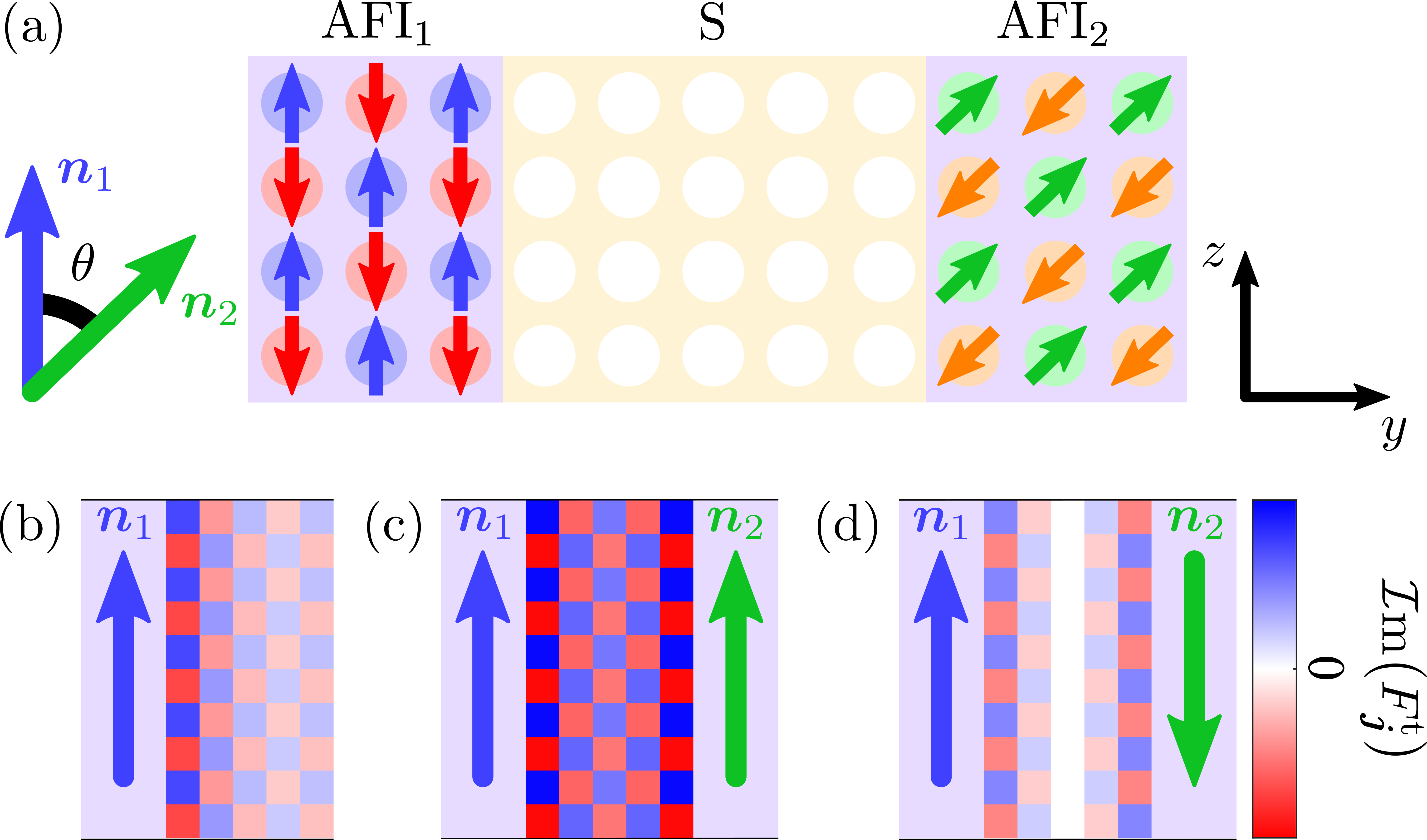}
	\caption{Schematic depiction of the system and key underlying phenomena. (a)~A conventional superconductor (S) is sandwiched between two \lina{compensated} antiferromagnetic insulators (AFIs) bearing N\'eel vectors $\bm{n}_1$ and $\bm{n}_2$ that subtend an angle $\theta$. (b)~In an \lina{AFI--S} bilayer, spatially alternating spin-splitting induced by the AFI predominantly generates N\'eel spin-triplet Cooper pairs characterized by a checkerboard pattern of their amplitude $F^{\text{t}}_{\ve{j}}$~\cite{Bobkov2022} \ak{thus manifesting an alternating spatial parity}. (c)~In an \lina{AFI--S--AFI} trilayer with odd number (considered 5 here) of S monolayers and $\theta = 0$, the N\'eel triplets generated by the two AFIs interfere constructively. This results in more induced spin-triplets and larger weakening of the spin-singlet superconductivity. (d)~If instead $\theta = \pi$, the N\'eel triplets from the two \lina{AFI--S} interfaces interfere destructively and superconductivity is weakened less. This dependence of the superconducting state on $\theta$ is reversed when the number of S monolayers is even due to the checkerboard pattern associated with the N\'eel triplets.}
	\label{fig:1}
\end{figure}

Intriguingly and subsequently, even a compensated interface with an AFI [Fig.~\ref{fig:1}(a)] was found to be spin-active~\cite{Johnsen2021}. This has recently been understood as being due to the generation of the so-called N\'eel triplet Cooper pairs~\cite{Bobkov2022}. These have been so named as their amplitude changes sign from one lattice site to the next [Fig.~\ref{fig:1}(b)], while the magnitude varies slower on the coherence length scale. This alternation of sign is due to the N\'eel triplets being formed from interband pairing~\cite{Bobkov2022}. Alternately, within an extended Brillouin zone scheme, they can be considered to result from finite-momentum pairing. In contrast, the regular spin-triplet Cooper pairs generated by an adjacent F only manifest a gradual spatial variation at the coherence length scale associated with the usual intraband pairing~\cite{Bergeret2005,Buzdin2005}.

In this Article, we theoretically investigate how such N\'eel triplet Cooper pairs enable intriguing phenomena in an \lina{AFI--S--AFI} trilayer [Fig.~\ref{fig:1}(a)]. Employing the Bogoliubov-de Gennes framework, we show that the \lina{critical temperature of the S layer} depends on the angle $\theta$ between the two AFIs' N\'eel vectors via a dominant $\cos \theta$ variation and a weaker $\sin^2 \theta$ contribution. The dominant effect is due to the constructive [Fig.~\ref{fig:1}(c)] or destructive [Fig.~\ref{fig:1}(d)] interference between the N\'eel triplets generated by the two \lina{AFI--S} interfaces~\cite{Bobkov2022}, while the $\sin^2 \theta$ term is reminiscent of equal-spin triplets resulting from noncollinearity between the two N\'eel vectors~\cite{Volkov2003,Fominov2010,Leksin2012,Chourasia2023}. For large enough \lina{AFI--S} interfacial exchange coupling, a complete \ak{and abrupt} suppression of superconductivity (i.e., $T_c \to 0$) is achieved. Due to the alternating sign of N\'eel triplet correlations, the $T_c$ vs.~$\theta$ dependence reverses when the number of S monolayers changes parity providing a \ak{distinct and unique} signature of the N\'eel triplets' role. By computing the superconducting free energy density as a function of $\theta$, we further demonstrate that the generated N\'eel Cooper pairs mediate coupling between the two AFIs' N\'eel vectors exhibiting the signature parity effect with the S monolayers number. Our theoretical results suggest a direct experimental probe of these recently predicted N\'eel triplets~\cite{Bobkov2022} while enabling antiferromagnetic superconducting spintronics devices.

%-----------------------------------------Theory------------------------------------------------------- %

\section{System and theoretical model}
We consider a thin-film superconductor which on each side is interfacing an antiferromagnetic insulator, as schematically depicted in Fig.~\ref{fig:1}(a). While electron hopping is only allowed within the S layer, the two AFIs impose a local spin-splitting via interfacial exchange onto the atomic layer closest to the \lina{S--AFI} interfaces~\cite{Kamra2018,Bobkov2022}. We can thus describe the system by the Hamiltonian
\begin{align}
    H =& 
    -t\sum_{\langle\ve{i},\ve{j}\rangle,\sigma}c_{\ve{i},\sigma}^{\dagger}c_{\ve{j},\sigma}
    -\mu\sum_{\ve{j},\sigma}c_{\ve{j},\sigma}^{\dagger}c_{\ve{j},\sigma}
    -\frac{J}{2}\sum_{\ve{j}}\ve{M}_{\ve{j}}\cdot\ve{S}_{\ve{j}}
    \notag\\
    &+\sum_{\ve{j}}\left(\frac{|\Delta_{\ve{j}}|^2}{U}+\Delta_{\ve{j}}^* c_{\ve{j},\downarrow}c_{\ve{j},\uparrow}+\Delta_{\ve{j}}c_{\ve{j},\uparrow}^{\dagger}c_{\ve{j},\downarrow}^{\dagger}\right).
    \label{eq:H}
\end{align}
Here, $c_{\ve{j},\sigma}^{(\dagger)}$ is the annihilation (creation) operator associated with an electron of spin $\sigma$ at lattice site $\ve{j} \equiv (j_z,j_y)$, $t$ parametrizes electron hopping between nearest neighbor sites within the S, $\ve{S}_{\ve{j}} \equiv \sum_{\sigma,\sigma'} c_{\ve{j},\sigma}^{\dagger} \ve{\sigma}_{\sigma,\sigma'} c_{\ve{j},\sigma'}$ \lina{is the spin operator for S electrons with $\ve{\sigma}$ as the vector of Pauli matrices}, and $\Delta_{\ve{j}} \equiv -U\langle c_{\ve{j},\downarrow}c_{\ve{j},\uparrow}\rangle$ is the self-consistently evaluated mean-field superconducting gap~\cite{Zhu2016}. The chemical potential $\mu$ is adjusted to fix the filling fraction, which we assume as $n=0.5$ here. We consider the S lattice to bear the size $N_z \times N_y$ with periodic boundary conditions along $\ve{z}$ [Fig.~\ref{fig:1}(a)]. \ak{As we consider ideal insulating antiferromagnets, their thicknesses do not influence the phenomena investigated here.}

A local spin-splitting field $J\ve{M}_{\ve{j}}/2$ is imposed by the two AFIs onto the S interfacial monolayers $(j_z,1)$ and $(j_z,N_y)$. Here, $J$ parametrizes the \lina{AFI--S} interfacial exchange coupling. As depicted in Fig.~\ref{fig:1}(a), the magnetic moments in the first AFI have a fixed orientation corresponding to the N\'{e}el vector $\ve{n}_1=\ve{z}$ so that $\ve{M}_{(j_z,1)}=(-1)^{j_z-1}\ve{n}_1$. The N\'{e}el vector $\ve{n}_2 =[\cos(\theta)\ve{z}+\sin(\theta)\ve{y}]$ of the second AFI leads to rotation of the local spin-splitting oriented along $\ve{M}_{(j_z,N_y)}=(-1)^{j_z-1}\ve{n}_2$. 

We numerically diagonalize the Hamiltonian in Eq.~\eqref{eq:H} by solving the Bogoliubov--de Gennes equation \cite{Zhu2016} self-consistently:
\begin{align}
	H = & H_0 + \sum^{'}_n E_{n}\gamma_n^{\dagger}\gamma_n \quad \mathrm{with} \label{eq:Hdiag} \\
	H_0 = & -N \mu - \sum_{\ve{j}} \frac{|\Delta_{\ve{j}}|^2}{U} - \frac{1}{2} \sum^{'}_n E_n, \label{eq:H0}
\end{align}
where $\sum^{'}_n$ denotes the sum over positive eigenenergies $E_n>0$ only, $\{\gamma_n^{\dagger}\}$ is a set of \lina{unique} fermion operators, and $N = N_z N_y$ is the total number of S lattice sites. The resulting solution provides complete information on the superconducting or normal state of the S layer.

%-----------------------Critical temperature dependence-----------------------$

\section{Critical temperature control via $\theta$}
In order to examine the magnetoresistance and S \lina{layer's} critical temperature dependence on the AFIs, we numerically compute the superconducting critical temperature $T_c$. It is determined using a binary search algorithm locating the temperature at which the superconducting gap starts to increase from a near-zero initial guess upon its self-consistent evaluation~\cite{Johnsen2019,Johnsen2020}.

\begin{figure}[tb]
	\centering
	\includegraphics[width=85mm]{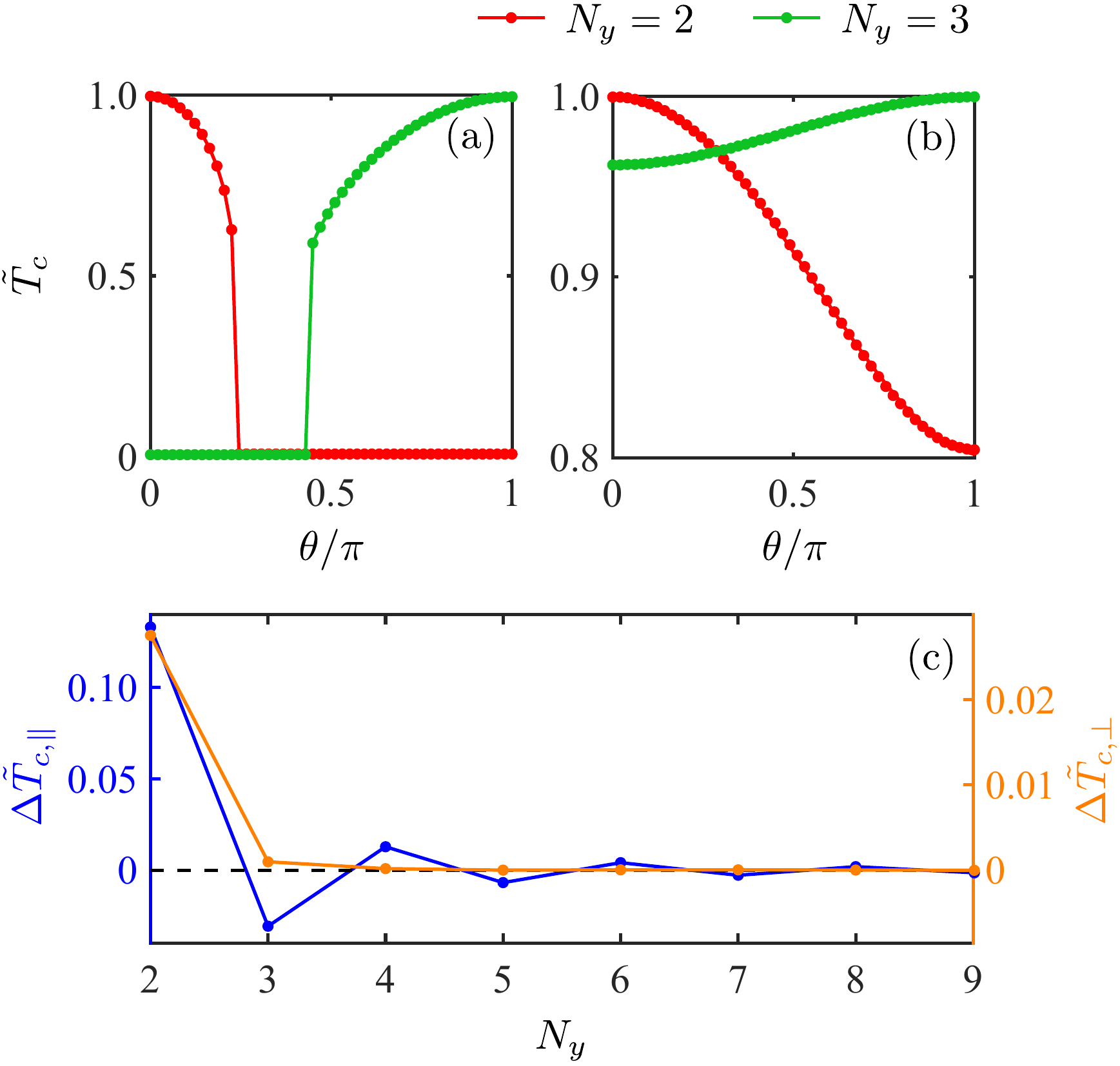}
	\caption{Normalized critical temperature $\tilde{T}_c$ variation with $\theta$ for (a) stronger and (b) weaker interfacial exchange coupling $J$. The variation is reversed when the number of S monolayers $N_y$ changes from \lina{even to odd}. A complete suppression of $T_c$ is observed for the stronger $J$ case (a), while the weaker exchange (b) results in variation of $T_c$ as per Eq.~\eqref{eq:Tcvstheta}. (c)~\lina{By f}itting the numerically evaluated $\tilde{T}_c(\theta)$ to Eq.~\eqref{eq:Tcvstheta} for different \ak{thicknesses} $N_y$, $\Delta \tilde{T}_{c,\parallel}$ and $\Delta \tilde{T}_{c,\perp}$ are obtained and \ak{studied for their thickness $N_y$ dependence}. The parity effect of $\Delta \tilde{T}_{c,\parallel}$ with respect to $N_y$ results from the alternating sign of N\'eel triplets' amplitude, as discussed in Fig.~\ref{fig:1}. In all panels, $N_z = 202$. In (a), $U/t = 1$ and $J/t = 0.08$. In (b), $U/t = 1$ and $J/t = 0.02$. In (c), $U/t = 1.3$ and $J/t = 0.08$.}
	\label{fig:2}
\end{figure} 

\ak{To succinctly capture and present the $T_c$ variation with $\theta$ for different thicknesses $N_y$ of the S layer, we first parametrize $T_c$ vs.~$\theta$ on symmetry grounds. This parametrization is only valid for small changes in $T_c$.} For a small $J$, $T_c$ is only weakly altered by the adjacent AFIs and \ak{is expected to bear} the dependence:
\begin{align}\label{eq:Tcvstheta}
	\tilde{T}_c(\theta) \equiv \frac{T_c(\theta)}{T_{c,0}} \equiv &~ \Delta \tilde{T}_{c,\parallel} \cos \theta + \Delta \tilde{T}_{c,\perp} \sin^2 \theta + \tilde{T}_{c,\parallel}, 
\end{align}
where $T_{c,0}$ is the critical temperature of the same S layer when it is not coupled to the AFIs, i.e., assuming $J = 0$ in Eq.~\eqref{eq:H}. From Eq.~\eqref{eq:Tcvstheta} above, we see \lina{that}
\begin{align}
	\Delta \tilde{T}_{c,\parallel} = & ~ [\tilde{T}_c(0) - \tilde{T}_c(\pi)]/2, ~~ \tilde{T}_{c,\parallel} = [\tilde{T}_c(0) + \tilde{T}_c(\pi)]/2, \nonumber \\
	\Delta \tilde{T}_{c,\perp} = & ~\tilde{T}_c(\pi/2) - \tilde{T}_{c,\parallel}.
\end{align}
In \lina{Eq.~\eqref{eq:Tcvstheta}}, the $\Delta \tilde{T}_{c,\parallel} \cos \theta$ term is \ak{expected due to} the interference of zero-spin N\'eel triplets generated by the two \lina{AFI--S} interfaces~\cite{Bobkov2022}, as briefly outlined in Fig.~\ref{fig:1}. \ak{It is analogous to the $\cos \theta$ dependence in \lina{F--S--F} trilayers~\cite{Degennes1966,Deutscher1969} and bears the symmetry of vectorial addition of the spin-splitting fields from the two AFIs.} The $\Delta \tilde{T}_{c,\perp} \sin^2 \theta$ term \ak{is expected} from the generation of equal-spin triplets via the noncollinearity between $\ve{n}_1$ and $\ve{n}_2$~\cite{Volkov2003,Fominov2010,Leksin2012,Chourasia2023} \ak{as it is finite only when the two magnetic orders are noncollinear}. \ak{In Eq.~\eqref{eq:Tcvstheta}, $\Delta \tilde{T}_{c,\parallel}$ characterizes the $T_c$ difference between parallel and antiparallel configurations. When it is positive (negative), the $T_c$ is larger for the parallel (antiparallel) configuration of the magnetic orders. On the other hand, $\Delta \tilde{T}_{c,\perp}$ represents the change in $T_c$ when going from parallel to perpendicular configurations. Together, $\Delta \tilde{T}_{c,\parallel}$ and  $\Delta \tilde{T}_{c,\perp}$ provide a succinct parametrization to study and present $T_c$ vs.~$\theta$ in our system. We emphasize that our numerical evaluation of $T_c$ does not depend on or assume this parametrization [Eq.~\eqref{eq:Tcvstheta}].}

In Fig.~\ref{fig:2}(a), we depict the $\tilde{T}_c$ variation when the interfacial exchange is strong and results in a complete $T_c$ suppression for certain $\theta$. When the number of S monolayers $N_y = 2$, the N\'eel triplets generated by the two \lina{AFI--S} interfaces interfere destructively for $\theta = 0$. This results in a weakening of the effect due to the AFIs and a larger $T_c$ at $\theta = 0$. For $N_y = 3$, the interference becomes constructive for $\theta = 0$ [Fig.~\ref{fig:1}(c)] due to the checkerboard pattern of the N\'eel triplets [Fig.~\ref{fig:1}(b)] and the $T_c$ vs.~$\theta$ trend is reversed. When the exchange coupling $J$ is small enough to avoid a complete suppression of $T_c$, \ak{the numerically evaluated $\tilde{T}_c(\theta)$ [Fig.~\ref{fig:2}(b)] is found to perfectly fit \lina{Eq.~\eqref{eq:Tcvstheta}}}. The reversal of trends between $N_y = 2$ and 3 remains as before and is attributed to the interference and checkerboard effects. 

\ak{Considering a filling fraction $n = 0.6$, we found a negligible dependence of $T_c$ on $\theta$. This is consistent with a much weaker generation of N\'eel triplets away from $n = 0.5$ corresponding to $\mu = 0$~\cite{Chourasia2023}. Furthermore, for a direct comparison, we discuss plots analogous to Figs.~\ref{fig:2}(a) and (b) for a trilayer comprising ferromagnetic insulator (FI) instead of AFI in the Appendix. The FI--S--FI trilayer is found to exhibit a weaker $T_c$ dependence, lack of an abrupt jump to 0 seen in Fig.~\ref{fig:2}(a), and no reversal of $T_c$ variation between $N_y = 2$ and $3$. This emphasizes the several unique features of our investigated AFI--S--AFI system. Here, we have considered AFIs with zero net magnetic moments. In the presence of a finite magnetic moment due to canting~\cite{Chourasia2023}, we expect the $T_c$ variation to bear a small contribution reminiscent of the FI--S--FI case investigated in the Appendix.}

Finally, Fig.~\ref{fig:3}(c) shows the dependence of $\Delta \tilde{T}_{c,\parallel}$ and $\Delta \tilde{T}_{c,\perp}$ on $N_y$ obtained by fitting the numerically evaluated data to Eq.~\eqref{eq:Tcvstheta}. $\Delta \tilde{T}_{c,\parallel}$, found to be an order of magnitude larger than $\Delta \tilde{T}_{c,\perp}$, exhibits a parity effect with $N_y$ due to the checkerboard pattern of N\'eel triplets [Fig.~\ref{fig:1}(b)] and the resulting interference effects [Figs.~\ref{fig:1}(c) and (d)]. This further validates the argument presented above that the $\Delta \tilde{T}_{c,\parallel} \cos \theta$ term stems from the N\'eel zero-spin triplets~\cite{Bobkov2022,Chourasia2023}. As $\Delta \tilde{T}_{c,\perp}$ stems from the regular equal-spin triplets generated by the noncollinearity between $\ve{n}_1$ and $\ve{n}_2$~\cite{Volkov2003,Fominov2010,Leksin2012}, it exhibits a simple decay with $N_y$ without any alternation of its sign.  

The results presented above \lina{(Fig.~\ref{fig:2})} show that an infinite magnetoresistance~\cite{Li2013}, resulting from a switching between the normal resistive and superconducting states using an applied magnetic field, is achievable in the considered \lina{AFI--S--AFI} trilayer by reorienting the N\'eel vector of one AFI with respect to the other. Recent experiments already demonstrate manipulation of the N\'eel vector in an easy-plane AFI, such as hematite above the Morin transition~\cite{Morrish1995}, using small magnetic fields~\cite{Wimmer2020}. Furthermore, a complete suppression of $T_c$ [Fig.~\ref{fig:1}(a)] enables such a device at arbitrarily low temperatures. An observation of the parity effect with $N_y$ \lina{[Fig.~\ref{fig:2}(c)]} will additionally provide evidence in favor of these recently predicted N\'eel triplets.

%-----------------------Triplets-mediated coupling-----------------------$

\section{N\'eel triplets-mediated coupling between the antiferromagnetic insulators}
\begin{figure}[tb]
    \centering
    \includegraphics[width=85mm]{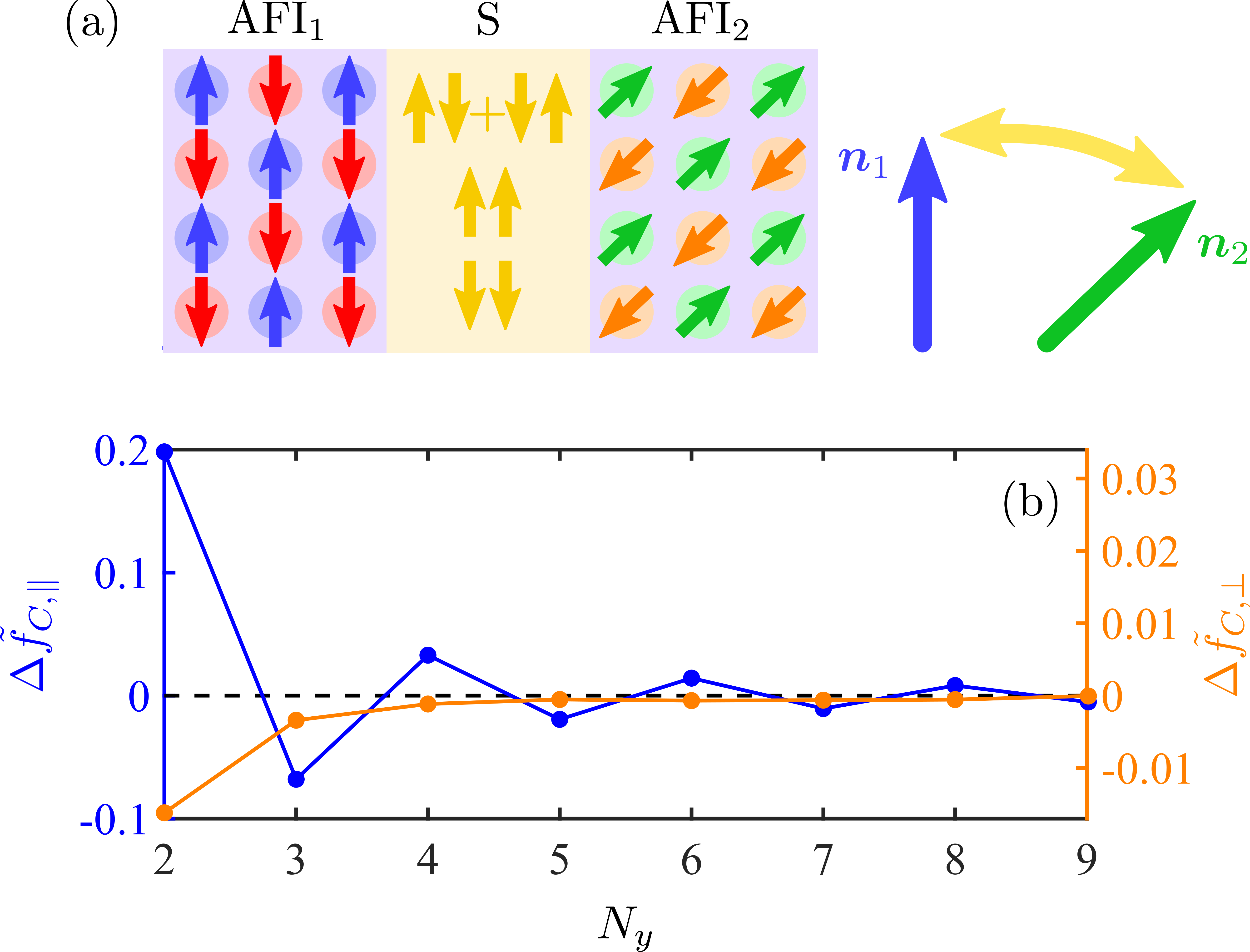}
    \caption{(a) Spin-triplet Cooper pairs generated by the AFIs mediate coupling between the two N\'eel vectors $\ve{n}_1$ and $\ve{n}_2$. This coupling is captured by the superconducting condensate's contribution $f_{C}$ to the $\theta$-dependence of the free energy density. (b) Fitting numerically evaluated $f_{C}$ to Eq.~\eqref{eq:fCvstheta} yields $\Delta \tilde{f}_{C,\parallel}$ and $\Delta \tilde{f}_{C,\perp}$, which are plotted vs.~$N_y$ \ak{thereby delineating the thickness dependence of the mediated coupling}. We have employed $N_z=202$, $U/t = 1.3$, $J/t = 0.05$, and $\beta t = 10^{4}$.}
    \label{fig:3}
\end{figure}

We have learned above how generation of N\'eel triplets by the two AFIs enables control over the superconducting state in an \lina{AFI--S--AFI} trilayer. Now, we seek to examine the inverse effect i.e., how the superconducting condensate enables a coupling between the two N\'eel vectors $\ve{n}_1$ and $\ve{n}_2$ [Fig.~\ref{fig:3}(a)]. \ak{This is distinct from the conventional exchange coupling between two magnetizations~\cite{Degennes1966}.} Since the latter vanishes in a typical AFI, coupling two AFIs is more challenging and rewarding. \ak{Furthermore, in an FI--S--FI trilayer, the coupling between the two magnetic orders mediated by Cooper pairs competes with a direct dipolar interaction between them. A lack of the latter in our AFI--S--AFI system makes the role of Cooper pairs more important.}

To examine the desired coupling, we need to compute the S \lina{layer} free energy density $f$ as a function of $\theta$. It is given by $f=-(1/\beta N)\text{ln}(Z)$, where the partition function $Z$ is $Z=\text{Tr}[\exp(-\beta H)]$. Here, \lina{$\beta = 1/k_B T$} with $k_B$ the Boltzmann constant and $T$ the temperature. Inserting the diagonalized Hamiltonian Eq.~\eqref{eq:Hdiag} into the free energy density expression above, we obtain
\begin{align}\label{eq:freedef}
	f = \frac{H_0}{N}-\frac{1}{\beta N}\sum_{n}^{'}\text{ln}\left(1+e^{-\beta E_n}\right),
\end{align}
which is evaluated numerically~\cite{Johnsen2021}. This free energy density [Eq.~\eqref{eq:freedef}] includes the contribution of quasiparticles, which can also mediate a coupling between the two AFIs~\cite{Nunez2006,Haney2007,Xu2008,Baltz2016}. Since we are interested in the superconducting condensate's role in mediating this coupling, we focus on the superconducting condensation energy density contribution: $f_{C} \equiv f_\mathrm{N} - f_\mathrm{S}$ attributed to the Cooper pairs~\cite{deGennes}. Here, $f_\mathrm{N}$ ($f_\mathrm{S}$) denotes the free energy density in the normal (superconducting) state and is obtained when $U = 0$ ($U \neq 0$) in the Hamiltonian Eq.~\eqref{eq:H}.

Making \ak{symmetry-based} and physical arguments similar to the ones put forward in assuming the $T_c$ dependence Eq.~\eqref{eq:Tcvstheta}, we \ak{expect} the relation
\begin{align}
	\tilde{f}_C(\theta) \equiv \frac{f_C(\theta)}{f_{C,0}} \equiv &~ \Delta \tilde{f}_{C,\parallel} \left( \ve{n}_1 \cdot \ve{n}_2 \right) + \nonumber \\
	& ~ \Delta \tilde{f}_{C,\perp} \left( \ve{n}_1 \times \ve{n}_2 \right)^2 + \tilde{f}_{c,\parallel}, \label{eq:fCvstheta}
\end{align}
where $f_{C,0}$ is the condensation energy for the same S layer without the adjacent AFIs i.e., considering $J = 0$ in Eq.~\eqref{eq:H}. Furthermore, we have expressed $\theta$ in terms of the N\'eel unit vectors to emphasize and clarify their mutual coupling [Fig.~\ref{fig:3}(a)].

Evaluating $\tilde{f}_{C}(\theta)$ numerically, we find the results to fit Eq.~\eqref{eq:fCvstheta} perfectly thereby \ak{vindicating it and} providing the desired $\Delta \tilde{f}_{C,\parallel}$ and $\Delta \tilde{f}_{C,\perp}$. These have been plotted in Fig.~\ref{fig:3}(b) for different number $N_y$ of S monolayers. $\Delta \tilde{f}_{C,\parallel}$ originates from the N\'eel zero-spin triplets~\cite{Bobkov2022,Chourasia2023} and captures an exchange-like interaction between the two AFIs' N\'eel orders. Consequently, it also bears the parity effect resulting from the alternating nature of the pairing amplitude [Fig.~\ref{fig:1}(b)]. On the other hand, $\Delta \tilde{f}_{C,\perp}$ represents an unconventional interaction originating from the equal-spin triplets induced by the noncollinearity between $\ve{n}_1$ and $\ve{n}_2$. 

Altogether, Fig.~\ref{fig:3} delineates the \ak{thickness dependence of the} desired coupling between the two AFIs' N\'eel orders, which may find applications in control over AFIs. At the same time, the similarity between Figs.~\ref{fig:2}(c) and \ref{fig:3}(b) indicates the complementarity between and a common origin of the effects investigated here providing valuable insights into the N\'eel proximity effect and Cooper pairs.

%--------------------------------Summary--------------------------------$

\section{Conclusion}
Employing the Bogoliubov-de Gennes framework and numerical diagonalization of the Hamiltonian, we have demonstrated a control of the superconducting critical temperature ($T_c$) and a condensate-mediated coupling between the two N\'eel vectors in an \lina{AFI--S--AFI} trilayer with compensated interfaces. \ak{Our investigated trilayer manifests various advantages over its conventional FI--S--FI counterpart including a stronger effect on $T_c$ and no interference from magnetostatic fields in the coupling between the two magnetic orders.} The demonstrated dependence of $T_c$ on the two N\'eel vectors enables an infinite magnetoresistance in the current-in-plane geometry via switching between the normal and superconducting states of the S layer~\cite{Li2013}. An interference between the spin-triplet Cooper pairs generated by the \lina{AFI--S} interfaces is further shown to enable coupling between the two AFIs' N\'eel vectors $\ve{n}_1$ and $\ve{n}_2$. The predominant coupling mediated by the N\'eel zero-spin triplets is exchange-like $\sim \ve{n}_1 \cdot \ve{n}_2$, while a weaker coupling of the form $\sim (\ve{n}_1 \times \ve{n}_2)^2$ is caused by equal-spin triplets. These phenomena are enabled by the recently predicted N\'eel triplet Cooper pairs~\cite{Bobkov2022} generated at such compensated \lina{AFI--S} interfaces. As a result they bear a distinct parity effect carrying signatures of the N\'eel triplets' alternating amplitude and should provide the means to experimentally observe them. \ak{Thus, our work paves the way for investigating a broad range of superconducting hybrids incorporating antiferromagnets including the effects of spin-orbit coupling.} At the same time, the phenomena discussed here outline possibilities for exploiting the broad range of advantages offered by antiferromagnets in superconducting spintronic devices and phenomena.     

%-----------------------------------------Acknowledgment------------------------------------------------------- %

\begin{acknowledgments}
S.C. and A.K. acknowledge financial support from the Spanish Ministry for Science and Innovation -- AEI Grant CEX2018-000805-M (through the ``Maria de Maeztu'' Programme for Units of Excellence in R\&D) and grant RYC2021-031063-I funded by MCIN/AEI/10.13039/501100011033 and ``European Union Next Generation EU/PRTR''. L.J.K. acknowledges financial support from the Research Council of Norway through its Centers of Excellence funding scheme, project 262633, ``QuSpin''. \lina{I.V.B., G.A.B. and V.M.G. acknowledge support from MIPT, Project FSMG-2023-0014.}
\end{acknowledgments}

\appendix

\section{Ferromagnet-superconductor-ferromagnet trilayer}\label{sec:FI}

\begin{figure}[tb]
	\centering
	\includegraphics[width=85mm]{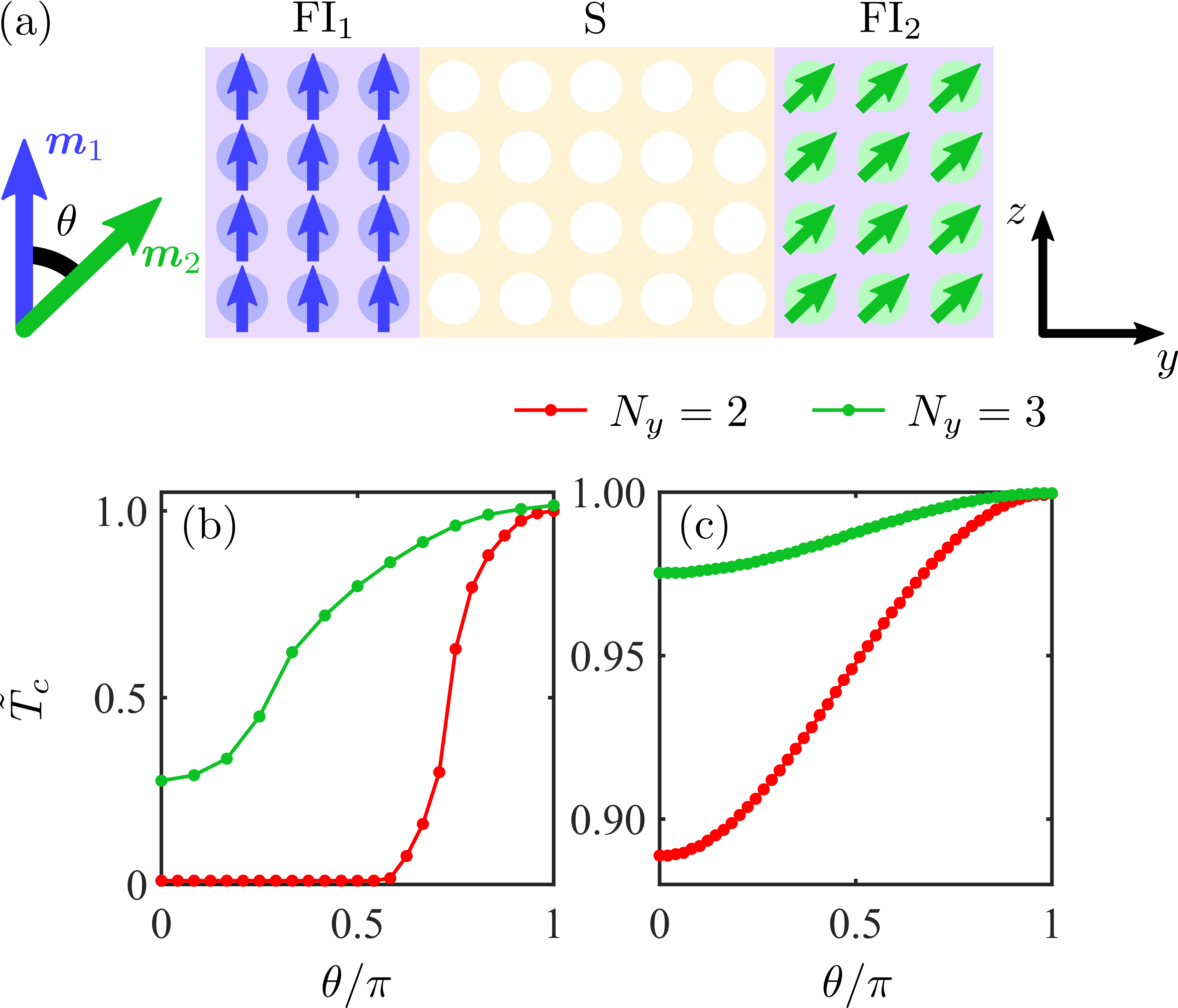}
	\caption{(a) Schematic depiction of a system where the compensated AFIs are replaced with ferromagnetic insulators (FIs) with magnetization along $\bm{m}_1$ and $\bm{m}_2$ that subtend an angle $\theta$. The normalized critical temperature $\tilde{T}_c$ variation with $\theta$ is plotted for (b) stronger and (c) weaker interfacial exchange coupling $J$. Contrary to the AFI/S/AFI system, the variation is no longer reversed when the number of S monolayers $N_y$ changes from even to odd. The weaker exchange (b) results in variation of $T_c$ as per Eq.~(4) in the main text. In both panels, $N_z = 402$ and $U/t = 1$. In (b), $J/t = 0.08$. In (c), $J/t = 0.02$.}
	\label{fig:SM1}
\end{figure}

Here, we consider trilayers comprised by ferromagnetic insulator (FI) layers [see Fig.~\ref{fig:SM1}(a)] instead of the compensated antiferromagnetic insulator (AFI) layers considered in the main text. Employing the same numerical routines, we evaluate $T_c$ as a function of the angle $\theta$ between magnetic orders of the two FI layers. The plots in Figs.~\ref{fig:SM1}(b) and (c) show data analogous to that in Figs.~2(a) and (b) of the main text, including use of the same parameter values. 

There are two minor differences in the numerical method though. First, we needed to employ a larger value of $N_z$ to adequately capture the superconducting properties. This is because the density of states for the AFI-S-AFI is larger than its FI-S-FI counterpart for the considered parameters. Thus, a larger number of lattice sites was needed to obtain convergent values for the FI-S-FI trilayer. Secondly, the algorithm for $T_c$ evaluation needs to be modified and in each binary search iteration, the superconducting gap needs to be established self-consistently. This is because the FI-S-FI system manifests a first-order phase transition with temperature~\cite{Chandrasekhar1962,Clogston1962} for larger values of $J$. There are thus multiple stable solutions for the superconducting state for a range of parameters. Our updated $T_c$ search algorithm overcomes all these complications. As a result, the numerical evaluation for the FI-S-FI system was much more computationally intensive, resulting in our providing fewer data points in Fig.~\ref{fig:SM1}(b). 

The numerically evaluated data plotted in Figs.~\ref{fig:SM1}(b) and (c) shows that the variation of $T_c$ is weaker in the case of FI-S-FI, as compared to the situation in AFI-S-AFI. Furthermore, the maximum $T_c$ is always obtained at $\theta = \pi$, as expected~\cite{Degennes1966}. There is no parity effect with the number of S layers, reinforcing our argument that the observed parity effect in AFI-S-AFI system is a smoking gun signature of the N\'eel triplets. Finally, Fig.~\ref{fig:SM1}(b) shows that for the FI-S-FI system, $T_c$ varies smoothly with $\theta$ and there is no abrupt jump to 0, like what is seen for the AFI-S-AFI case.

\bibliography{AF_S_AF}

\end{document}